\documentclass[acus]{JAC2000}

\usepackage{graphicx}

\begin{document}
\title{\flushright{PSN:WEAP030}\\[15pt] \centering EXPERIENCE OF USING
MULTIMONITOR WORKSTATIONS UNDER XFREE86 4.X IN VEPP-5 CONTROL ROOM}

\author{D.Yu.Bolkhovityanov, R.G.Gromov, I.L.Pivovarov, A.A.Starostenko\\
The Budker Institute of Nuclear Physics, Novosibirsk, Russia}

\maketitle

\begin{abstract}
Modern PC workstations often provide more CPU power than required for most
control applications. On the other hand, the screen space is always in short
supply. One possible solution is to use more PCs, but in fact we need only
more screens, not more keyboards, mice etc. PC architecture allows using more
than one videocard, and X Window protocol is also aware that there can be more
than one screen. But until release of XFree86 version 4 there was no freely
available server capable of driving multiple ``heads''. We have been using
multiheaded workstations under XFree86 in the VEPP-5 control room since early
2000 (currently 4 4-headed PCs plus several dual-headed). The ``Xinerama''
mode (one-large-screen) is better suited for accelerator control system than
``several separate screens''. When moving to this configuration we've
encountered a number of, mainly human-related, problems, some of which
required modifications to X server Additionaly, the ``style'' of performing
control has slightly changed.
\end{abstract}

\section{NEED IN MORE SCREEN SPACE}

Historically automation at BINP is based on CAMAC.  A home-made Odrenok
\cite{odrenok} machines were used as both crate controllers and as the main
computational power.  The information was displayed via CAMAC-based display
controllers, which gave color 256$\times$256 pixel picture.  That allowed
sufficient display space for most tasks.

On the new VEPP-5 facility the computation and high-level control was moved
from crate controllers to Intel-based workstations.  So, the aging CAMAC
display hardware wasn't an option.

Modern video cards and monitors have resolutions large enough to simply put
contents of all 256$\times$256 displays on them.

This approach was taken by the VEPP-4 team, which exploits a large number of
legacy programs using CAMAC display controllers.  They made an emulation
library, which redirects graphic output of such programs to X11 windows.  But
there was no reason for VEPP-5 to go this way.

\section{POSSIBILITY}

PCI bus allows to have multiple videocards in one computer.  One card is
treated as primary (the one on which the boot screen appears), and others are
inactive until a multihead-aware system is loaded.  AGP slot looks like just
one more PCI slot.

From the very beginning X theoretically allowed to use several screens on one
host.  These screens are referred to as {\it{}hostname:N.0},
{\it{}hostname:N.1}, etc., where {\it{}N} after semicolon is a display number
(typically {\it{}0}) and {\it{}0}, {\it{}1}, etc. after dot is a screen
number.

But in practice, XFree86 up to version 3.x inclusive didn't support multihead.
That capability appeared in long-awaited version 4.0, released in early 2000.

\section{TRADITIONAL MULTIHEAD VS XINERAMA}

\begin{figure}[htbp]
\includegraphics[width=\linewidth]{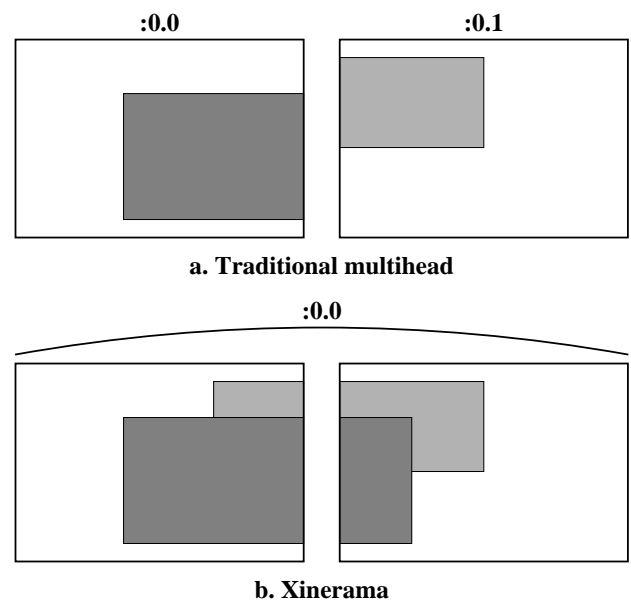}
\caption{Traditional and Xinerama multihead}
\label{f:xinerama}
\end{figure}

The traditional X multihead presents each screen separately, so that when a
window is created, it is placed on one of these screens and cannot span
screens or be moved from one screen to another (see Fig.\ref{f:xinerama},a).

On the other hand, Xinerama makes multiple physical screens behave as a single
screen, transparently to the clients (see Fig.\ref{f:xinerama},b).  So, the
windows can be freely moved between screens.

A good source of information about Xinerama is \cite{XINERAMAHOWTO}.

Since the situation on the screen of control computer isn't static (there's
often a need to group windows in different ways, to move more important
windows to ``more visible'' screen), Xinerama is much better suited for use in
a control room than traditional multihead.

\section{XINERAMA PROBLEMS}

\subsection{Technical problems}

When joining screens, Xinerama leaves only depths which are common to all
screens.  So, it is impossible to join 16-bit screen with a 24-bit one. 
Additionally, the 24+8 ``overlay'' feature of Matrox cards is lost, since 2nd
head doesn't support it.

But the main inconvenience is that since all screens look like a single one to
all clients, the window managers happily place windows between screens,
maximize them on all screens, etc.

We use FVWM \cite{FVWM} in the VEPP-5 control room, so we invested some time
to its initial xineramification, which was completed by FVWM team (now
Xinerama support in FVWM is probably the most complete and configurable among
all WMs). Currently, most WMs are Xinerama-aware, but some toolkits still
aren't.\footnote{A frequent case: a window with Yes and No buttons is
centered, so that [Yes] goes to one monitor and [No] to the other one.}

\begin{figure}[htbp]
\includegraphics[width=\linewidth]{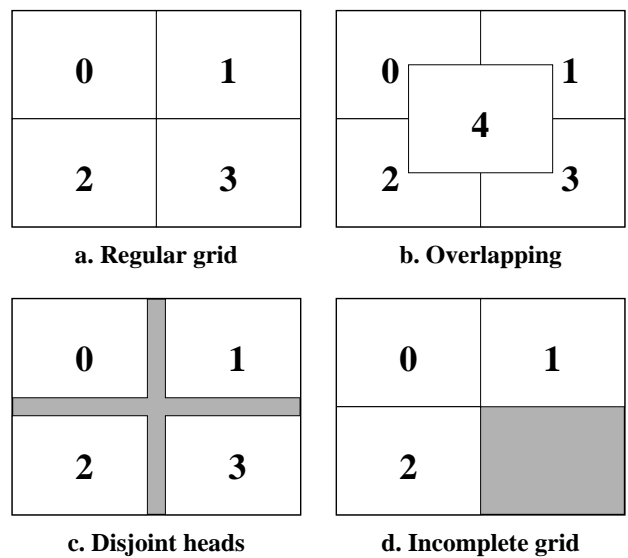}
\caption{Possible Xinerama layouts}
\label{f:layouts}
\end{figure}

And there is one more exotic problem.  Technically, Xinerama makes a single
large desktop with a size of a bounding rectangle of all screens.  And screens
themselves function as viewports to the desktop.  So, Xinerama allows to place
heads in many different ways (see Fig.\ref{f:layouts}).

\begin{enumerate}

\item[a] Heads can form a regular grid -- that's the most common case.

\item[b] They can overlap (a so-called ``clone/zoom'' mode).  This is used
very rarely, since the position of ``zoom'' screen is fixed and can't be moved
e.g. following the pointer.

\item[c] Can be disjoint\footnote{That's a pathological case; more often
screens of different sizes are used (e.g. 1024$\times$768 and 800$\times$600),
which has the same effect.}.

\item[d] Or the grid can be incomplete.

\end{enumerate}

In the two latter cases, there are ``black holes''
on the desktop, which aren't visible on any monitor and which can't be reached
with mouse.  The consequences are worst in the last case -- even complete
windows can disappear in the black hole.

Additionally, there are still some problems with software which either
requires a direct access to framebuffer, or uses a fullscreen mode (various
video capturing and movie playing programs).\footnote{Up to XFree86 4.1 Xawtv
behaved very funny: the window frame could be on one screen, and undecorated
video picture -- on another.}  But, thanks to XVideo extension, these problems
became very rare.

\subsection{Human Problems}

Some of our software developers are greedy: when they see so much display
space, they say: ``Hey, let's move this and this to another screen, and my
program will just fill this screen''.  The common rule is ``some programs tend
to grow to occupy all screen space''.  So, appetittes of some people need
reduction.

Another problem is that mouse pointer often gets lost on large screen space. 
Finally a patch for X server was developed \cite{visxcursor}, which allows to
1)~doublesize the pointer and/or 2)~change the default colors from
black\&white to something more visible.  We use red doublesized pointers,
which provides good visibility.

\section{HOW MANY HEADS TO USE}

The most common multihead layout in the world is two heads: side by side
horizontally, or one above another (if 2nd head is used rarely).

Three heads are hard to use: the layout will be either as on
Fig\ref{f:layouts},d, which is inconvenient, or lined up.  In the latter case
it takes too much time to move the pointer between the first and the last
screens.  When we {\it had} to use three heads, we put monitors in the shape
of ``r'' but X layout was ``three heads vertically''.  That setup was
extremely confusing for operators.

Four heads give the best balance between ``as much screen space as possible''
principle and convenience of use.  When used in a 2$\times$2 grid, as on
Fig.\ref{f:layouts},a, there are no black holes and the distance between heads
is small.

\section{HARDWARE}

\subsection{Criteria for selecting video cards}

First, that hardware should be multihead-capable (e.g., 3Dfx cards are known
not to work in multihead mode under XFree86 at all).  Second, it must have
good support in XFree86 and work very stable.  Third, it should produce an
excellent picture and have good 2D performance (3D isn't important).  Fourth,
video hardware should occupy as little PCI slots as possible.

\subsection{Solution we use}

There were 4 main manufacturers: ATi, Matrox, nVidia and S3 (the latter is
almost dead now).  We chose Matrox, because 1)~it satisfied all our criteria;
2)~we already had very positive experience with their products.

As to other two brands: nVidia doesn't open specifications of their cards, so that
XFree86 driver is very lacking, but instead nVidia provides binary-only
driver; ATi cards are not-so-good and there are myriads of subversions, which
affects stability of the driver.

We use MilleniumIIs as PCI cards, but any PCI card with 4M or more
memory\footnote{1152$\times$864~@~32bpp~$\approx$~4M RAM for framebuffer.}
would do (Millenium, G100, G200).

\subsection{Multiheaded videocards}

\begin{figure}[htbp]
\includegraphics[width=\linewidth]{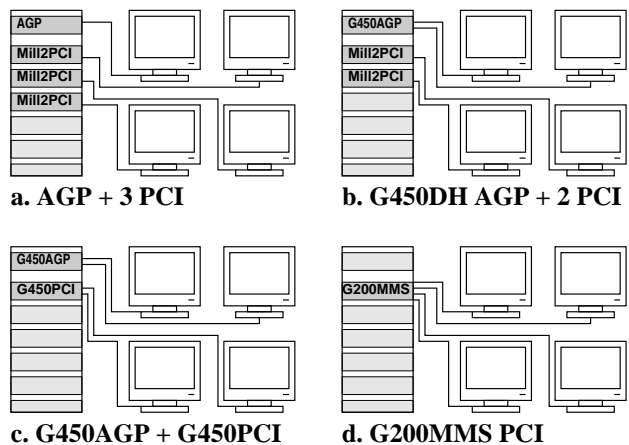}
\caption{Hardware options for 4 heads}
\label{f:hw}
\end{figure}

One more advantage of Matrox cards is that since 1999 they have two heads on
one card (G400DH, G450, G550).  So, to have 4 heads, an AGP G450 plus two PCI
MilleniumIIs were enough (see Fig.~\ref{f:hw},b), thus using only two PCI slots
(which are always in deficite in control machines).  

And Matrox produces a PCI version of G450, thus allowing to use only one PCI
slot in addition to AGP (see Fig.~\ref{f:hw},c).  Unfortunately, when running
as non-primary card, the G450 (either AGP or PCI) requires an additional
driver module, which is available as binary-only from Matrox (so-called HAL
module \cite{HAL}).  But we hope the native XFree86 support will become
better soon, thus making HAL redundant.

Currently two of our 4-headed PCs are equipped with
G450AGP+2$\times$MilleniumII, and two are G450AGP+G450PCI.

Theoretically there exists even a better choice -- G200MMS, which supports 4
heads on one card (see Fig.~\ref{f:hw},d), but it exists in PCI version only,
so if we need 4 heads total, it occupies the same one PCI slot as
G400AGP+G450PCI.  Additionally, G200MMS is almost impossible to find in
Russia.

\subsection{Motherboards}

We chose ASUS P3B-F (Intel 440BX chipset), which has 1 AGP slot, 6 PCI and one
ISA (one position is shared, so you get either 6 PCI and 0 ISA or 5 PCI and 1
ISA).  The main requirement was a presence of ISA slot, since we still use old
ISA hardware.

\section{FUTURE ENHANCEMENTS}

One feature our operators wish to have is an ability to control programs on
adjacent computers with their mouse.  A program {\it{}x2x} \cite{x2x} exists
which does exactly this, but it doesn't work with Xinerama.  So, we plan to
``xineramify'' {\it{}x2x}.

Currently the world moves to using TFT monitors, as those are more safe for
people.  But most TFTs have a limited viewing angle, which is inappropriate in
multiheaded system, and those TFTs which are okay (like SGI 1600SW) are too
expensive.  So, currently we use 17" CRT displays, but plan to replace them
when affordable TFTs appear.

Finally, consistent and ergonomic placement of windows on a 4-monitor desktop
is a time-consuming task.  So, we are planning to implement some sort of
automation for this.  Currently we are experimenting with X resource database
(the {\it{}WINDOWNAME.geometry} resource).

\end{document}